\documentclass[namedreferences]{solarphysics}
\usepackage{amsfonts}
\usepackage{mathtools}
\usepackage{multirow}
\usepackage[hyperref,optionalrh,solaromanenum]{spr-sola-addons} 
\usepackage{graphicx}                    
\usepackage{color}                       
\usepackage{breakurl}                         

\begin{document}

\begin{article}

\begin{opening}

\title{Solar flare catalog from 3 years of Chandrayaan-2 XSM observations}

%
\author[addressref={aff1,aff2}, corref]{\inits{A.B.V.}\fnm{Aravind Bharathi}~\lnm{Valluvan}}
\author[addressref={aff3}]{\fnm{Ashwin}~\lnm{Goyal}}
\author[addressref={aff3,aff4}]{\fnm{Devansh}~\lnm{Jain}}
\author[addressref={aff6,aff7}]{\fnm{Abhinna Sundar}~\lnm{Samantaray}}
\author[addressref={aff8,aff9}]{\fnm{Abhilash}~\lnm{Sarwade}}
\author[addressref={aff8,aff10}]{\fnm{Kasiviswanathan}~\lnm{Sankarasubramanian}}

%
\runningauthor{Valluvan et al.}
\runningtitle{XSM Solar Flare Catalog}

\address[id={aff1}]{Department of Physics, Indian Institute of Technology Bombay, Mumbai, 400076, India}
\address[id={aff2}]{Department of Astronomy and Astrophysics, University of California San Diego, SERF Building, La Jolla, CA 92093, USA}
\address[id={aff3}]{Department of Computer Science and Engineering, Indian Institute of Technology Bombay, Mumbai, 400076, India}
\address[id={aff4}]{Department of Computer Science, University of Illinois Urbana-Champaign, 201 N Goodwin Ave, Urbana, IL 61801, USA}
\address[id={aff6}]{Astronomisches Rechen-Institut, Heidelberg Universität, Mönchhofstraße 12-14, 69120 Heidelberg, Germany}
\address[id={aff7}]{Max-Planck-Institut für Astronomie, Königstuhl 17, 69117 Heidelberg, Germany}
\address[id={aff8}]{UR Rao Satellite Centre, Indian Space Research Organisation, Old Airport Road, Vimanapura PO Bengaluru, 560017, India}
\address[id={aff9}]{Department of Physics, Indian Institute of Technology Guwahati, 781039, India}
\address[id={aff10}]{Center of Excellence in Space Sciences India, Indian Institute of Science Education and Research Kolkata, Kolkata, 741246, India}

\begin{abstract}
We present a catalog of 6266 solar flares detected by the X-Ray Solar Monitor onboard the Chandrayaan-2 lunar orbiter between 1.55 and 12.4 keV (1 and 8 \AA) from 2019 September 12 to 2022 November 4, including 1469 type A flares. The catalog represents the first large sample, including both type A, hot thermal flares, and type B, impulsive flares, with a sub-A class sensitive instrument. We also detect 213 sub-A and 1330 A class flares. Individual flares are fit with an exponentially-modified Gaussian function and multi-flare groups are decomposed into individual flares. We validate our findings with flare catalogs made using visual inspection as well as automatic pipelines on Geostationary Operational Environmental Satellite and Solar Dynamics Observatory data. We find a clear bimodality in the ratio of the width to decay time between type A and B flares. We infer a power-law index of $\alpha_F = 1.92 \pm 0.09$ for the background-subtracted peak flux distribution of XSM flares, which is consistent with the value $\sim 2$ reported in the literature. We also infer $\alpha_F = 1.90 \pm 0.09$ for type B, and $\alpha_F = 1.94 \pm 0.08$ for type A flares, which has previously not been reported in the literature. These comparable values hint at a similarity in their generative processes. 
\end{abstract}

%

\end{opening}

%

\section{Introduction} \label{sec:intro}
Solar flares are stochastic, broadband emissions initiated in the Sun's corona, which span across the electromagnetic spectrum \citep{ackermann2014fermi,benz2017review,feng2020space}. They are associated with an increase in coronal plasma temperature \citep{ryan2012thermal}, with the largest flares producing temperatures in the tens of million kelvin \citep{sakurai2017heating,casadei2017anisotropy,jeffrey2018turbulence} and are often accompanied by coronal mass ejections and solar energetic particle events \citep{harrison1995nature, reames2021solar}. Solar flares are also known to have disruptive effects on the Earth’s magnetosphere \citep{tsurutani2009brief,lingam2017superflares} and form a component in holistic space weather monitoring \citep{feng2020space}. Although there is no broad consensus on their production mechanism, solar flares are hypothesised to be the result of magnetic reconnection in the solar corona \citep{klimchuk2006solving, shibata2011solar}. 

Solar physicists have been continuously observing the Sun for decades to understand the various energetics, including temperature and emission measures, as well as elemental abundance associated with solar flares \citep{feldman1992elemental,kepa2018analysis,mondal2021evolution}. The large temperatures attained in the Sun's corona remain an unsolved problem with results indicating that the largest flares only account for a fraction of the coronal heating. Thus, an abundance of micro- and nanoflares have been hypothesised to explain the remaining energy requirement \citep{parker1988nanoflares, hudson1991solar,sakurai2017heating}. Moroever, studies such as \cite{vadawale2021quietsun, nama2023coronal} have been limited by the need to visually inspect large datasets. Hence, there is a need for a large-scale, automatic pipeline to detect weak solar flares. Such a pipeline should also exploit the increased sensitivity in future solar missions carrying X-ray payloads such as Aditya-L1 \citep{sankarasubramanian2017}, and support the detection of the various solar flare types. 

Based on the time profile, spectrum and morphology of solar flares, three types of flares were proposed by \cite{tanaka1983types} and further investigated by \cite{tsuneta1987impulsive, dennis1988solar}. Type A or hot thermal flares show a gradual rise and fall arising from a single source, Type B or impulsive flares are the the result of a double source structure and show a rapid rise with an exponential decay, while Type C or coronal flares show no impulsive variation and have a strong, associated component in the microwave regime. Past algorithm implementations such as \cite{aschwanden2012_37years,goodman2019goes} have detected well-characterised, isolated type B flares at a consistent rate but their designs intrinsically lack the ability to detect other types of flares, type B flares {with low peak intensity} and individual flares in multi-flare groups. This is because the above implementations calculate and flag a rapid increase in observed flux as a flare and assume that the flux value will return close to the background level before the occurence of a subsequent flare. These impose limitations to their derived flare catalogs and has made it difficult to ascribe a precise definition to solar flares based on their X-ray light curve morphology alone. The effects of using an imprecise definition have been explored in \cite{ryan2016effects}, while there may be interesting implications for the waiting time distribution of solar flares upon decomposition of multi-flare groups \citep{wheatland2002understanding,aschwanden2021solarmemory}.

The paper is organised as follows. In Section \ref{sec:methods}, we provide an overview of the instrument and observation method. In Section \ref{sec:algorithm}, we describe our algorithm, the definition of a flare used in this study and the use of reliable filters once the flare catalog is generated. In Section \ref{sec:results}, we present the results of our large-scale, statistical study of solar flares, including the detection of type A flares and validation against existing flare catalogs. In Section \ref{sec:discussions}, we discuss these results in the context of other solar flare catalogs along with the limitations of this study. We present our conclusions in Section \ref{sec:conclusion}.

\section{XSM Observation Overview} \label{sec:methods}
The Solar X-ray Monitor (XSM) is a state-of-the-art X-ray payload on-board the Chandrayaan-2 (CH2) orbiter to monitor solar activity in the soft X-ray regime \citep{vadawale2014xsm, mithun2020xsmperf}. It samples the number of photons falling on the detector in the 0.8 to 15 keV bandwidth in one-second intervals \citep{shanmugam2015new,shanmugam2015radiation}. Calibrated, preprocessed light curves and spectra, referred to as level 2 files, with photon counts at one-second cadence are available on the Indian Space Science Data Centre archive, accessible via PRADAN\footnote{\url{https://pradan.issdc.gov.in/ch2/}}, in \texttt{FITS} format \citep{mithun2021data}. However, in this study, we have employed X-ray light curves over a smaller bandwidth from $1.55$ to $12.4$ keV ($1$ to $8$ \AA ), which have been calculated using pulse invariant spectral data. This will allow us to make direct comparisons with the pipelines developed for the Solar X-Ray Sensor (XRS) onboard the Geostationary Operational Environmental Satellite (GOES) \citep{goodman2019goes}. We restrict our analyses to 10 s binned data but this algorithm may also be run over one-second cadence data.

\begin{figure*}
    \centering
    \includegraphics[width=\linewidth]{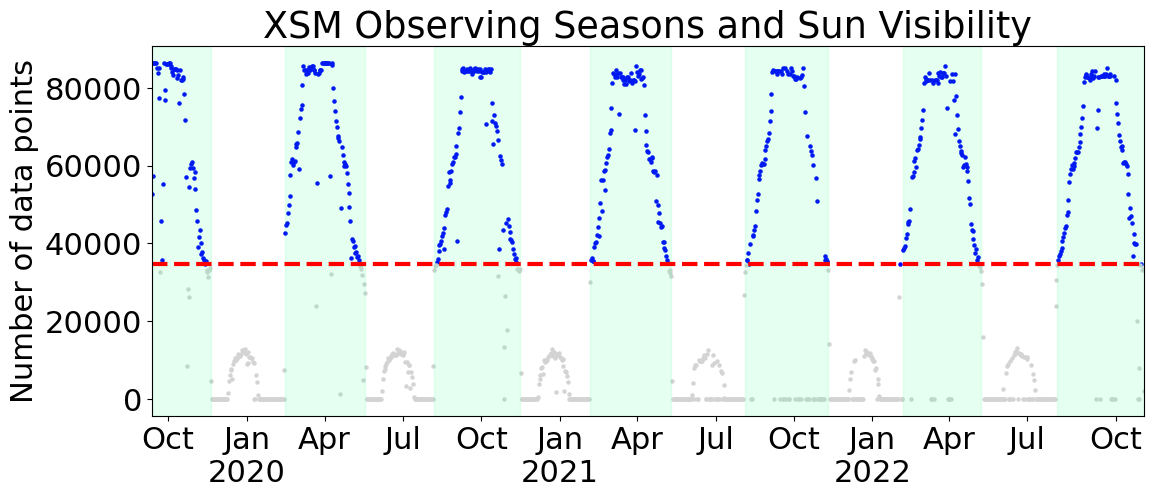}
    \caption{Seasonal variations of the Sun's visibility on XSM. Regions in pale green represent Dawn-Dusk season (D-D) and regions in white represent Noon-Midnight season (N-M). The dashed red line marks the 40\% level for the number of visible data points. Only the days in blue have been used to generate our XSM/CH2 solar flare catalog.}
    \label{fig:sunvisibility}
\end{figure*}

Observations using the XSM instrument started on September 12, 2019. As the satellite is on a lunar orbit, XSM does not always point towards the Sun. There are alternating three month periods of Dawn-Dusk season (D-D) when there is very good coverage of the Sun followed by three months of Noon-Midnight season (N-M) with sparse to no observations of the Sun. Readers are directed to \cite{vadawale2014xsm} for the full orbital specifications. In our analyses, we reject days with data points fewer than 40\% temporal coverage as suggested in \cite{mithun2020xsmperf}. The dates included in our analysis is captured in pale green in Figure \ref{fig:sunvisibility}. Note that there are a few days from May 2021 to the present for which the level 2 files are not available on the PRADAN website.

\section{Algorithm Description}\label{sec:algorithm}

\subsection{Background Estimation}

To automatically identify solar flares, we need to view flare flux relative to the background flux of the Sun at that time \citep{battaglia2005size,hannah2008rhessi}. This is because the background solar activities that constantly emit X-ray photons and contribute to the intensity, are not always due to flares \citep{usoskin2017history}. Therefore, to get accurate results, any analysis on these light curves need to be carried out after subtracting the background intensity \citep{blanton2011sdss}. 

Unlike astronomical transient searches for fast radio bursts, gamma ray bursts or photometric microlensing events, where the event rates are in the order of $\mathcal O(1)$ per day, the solar background intensity is tangled with the intermittent production of solar flares. Moreover, the arrival of an active region within the instrument's field of view rapidly affects the background intensity. 

We define a `flaring duration' as the part of the light curve which contains flares and should not be considered as a part of the background. Thus, for a good estimate of the background intensity, we need to know the non-flaring duration which requires removing the flaring duration and hence the detection of said flaring duration. Detection of flaring duration, in turn, requires a good background estimate leading to a circular problem. To circumvent this, we start with an initial guess for flaring durations and then follow through to get an estimate of the background. As the focus in this step is to estimate the background, it is acceptable for the algorithm to classify the flaring durations as larger than they actually are, i.e., overestimate the number of flaring durations in the day. 

We employ a 2-step process to smoothen the input light curve: mean-binning with bin size $= 120$ s followed by a convolution with a Gaussian kernel of standard deviation $\sigma_{G}$. This removes random noise in the light curve and improves the detection efficiency of solar flares \citep{mithun2020xsmperf, mithun2021xsmcalib}. It is to be noted that the smallest detectable (in terms of peak flux), sub-A class flares have flare durations in the order of a few minutes and will, therefore, not be lost to binning \citep{temmer2001statistical}. Furthermore, the Gaussian kernel reduces high frequency components which do not correspond to flares and which would have been detrimental to our algorithm's accuracy \citep{davenport2015detecting}. However, the endpoints of a time-series data are more drastically affected by convolution operations. Thus, we stitch light curves from the previous day and the next day to circumvent this issue, and then unstitch these light curves once the background is estimated. The default values of these parameters have been summarised in Table \ref{tab:consetting}. 

We find a preliminary list of peaks in the light curve using the \texttt{scipy.signal.} \texttt{find\_peaks} module (hereafter referred to as \texttt{scpeaks}){, in \texttt{SciPy v1.9.3} \citep{2020SciPy-NMeth}}. This method defines the topographic prominence as the height difference between a peak and the lowest contour line that completely encircles a given peak, with no higher peak being contained within it \citep{helman2005prominence}. We filter out peaks with topographic prominence higher than a threshold $\tau_{P}$ and refer to the list as initial flare peaks. 

We approximate the start time and end time of all the flaring durations, for which the peaks have already been found. In the following discussion, this is referred to as the `slope algorithm'. The slope algorithm is based on  \cite{aschwanden2012_37years}
for detecting impulsive flares. We start by choosing one of the peaks found by \texttt{scpeaks} and `walk down' one side until the magnitude of the slope, calculated using four consecutive data points in the binned and smoothened light curve, drops under a threshold value $\tau_{m}$. When this happens, we claim that the flaring duration has been constrained on that side. Depending on which side, i.e., left or right, we can approximate a start time and an end time. In effect, we are referring to the flares identified by \cite{aschwanden2012_37years} as flaring durations, approaching their results with caution. 

We recognise that one cannot reliably claim that the minima in intensity between two flaring durations found by this method will reach the background level, as the flaring durations may be too close to each other keeping the intensity above background throughout. Hence, we follow this up by merging consecutive flaring durations into a single flaring duration if the end of the first and the start of the second are closer than a threshold value $\tau_{merge}$.

\subsection{Detection of Flare Groups}

\begin{figure}
    \centering
    \includegraphics[width = 0.6\linewidth]{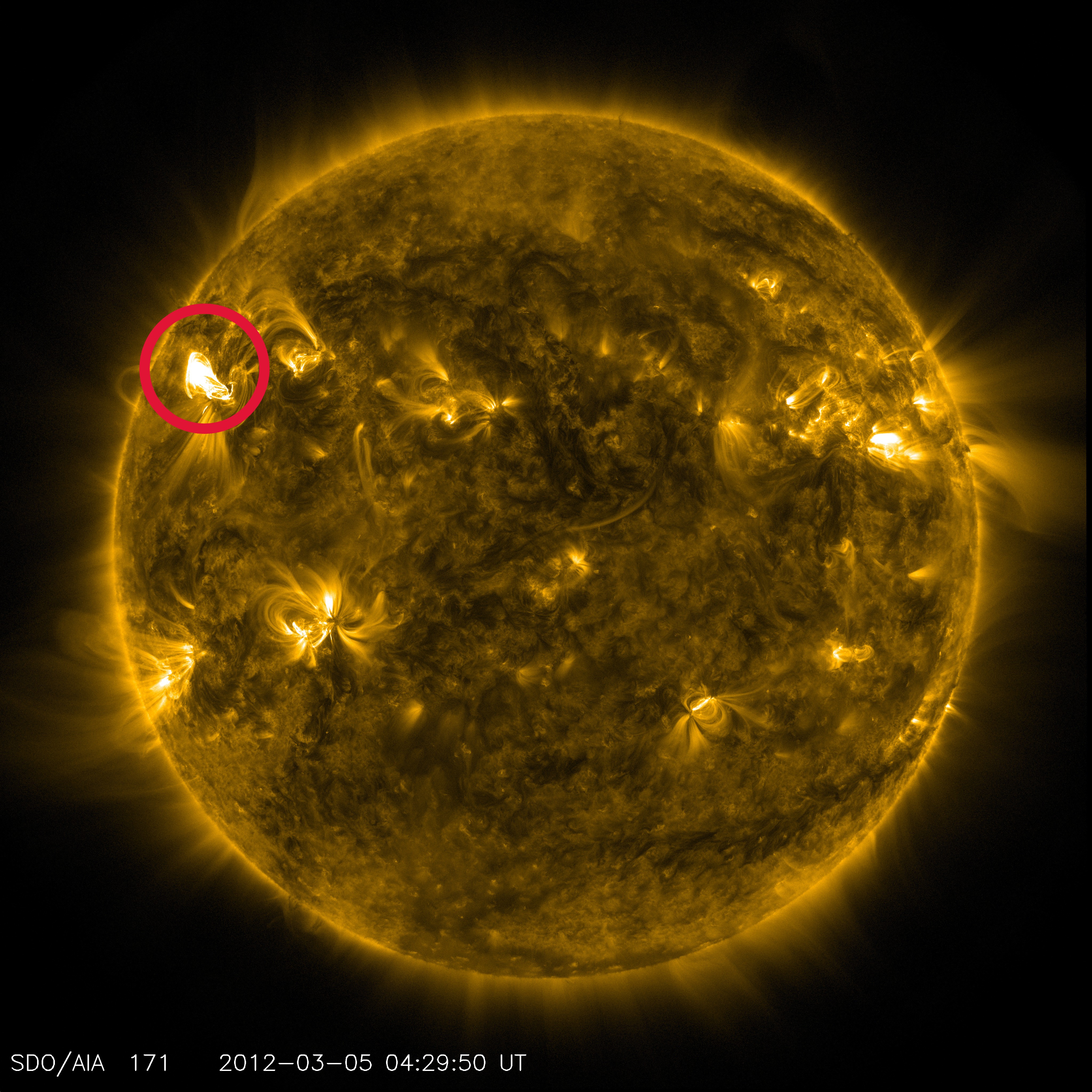}
    \caption{Integrated flux measurements from the 171 \AA\  channel (72.5eV) of SDO/AIA \citep{lemen2012aia}. The region encircled in red shows a flaring location on the Sun that emitted two overlapping flares at 2012-03-05 02:14 UT. Image credit: NASA/SDO/AIA.}
    \label{fig:aia171}
\end{figure}

`Flare groups' refer to single flares or overlapping multi-flares detected by the algorithm described in this section. Since XSM does not obtain spatial information, we cannot distinguish between multi-flares originating from the same flaring location (with a slight time offset) versus the multi-flares arising from different flaring locations on the Sun. Figure \ref{fig:aia171} shows multi-flare observations from the same as well as distinct flaring locations. We employ a flare decomposition step to distinguish these as individual flares. 

We use dynamic thresholding to detect potential flare groups based on the variability of the estimated background. We calculate the standard deviation $\sigma_{BG}$ of the estimated background and search for points exceeding the threshold $\tau_{FG} = n\sigma_{BG}$, where $n$ is a tunable parameter{ set by trial-and-error}\footnote{Although similar to a single iteration of sigma clipping, the value of $n$ here is lower than the typical 3 or 5-sigma-clip values.}. {While the variability in the solar background is remarkably low during the Quiet Sun observation periods, the background activity is high on active days \citep{vadawale2021quietsun} By employing a dynamic threshold, we can guard the algorithm against detecting false positive flares on active days of the Sun while maintaining a high accuracy on Quiet Sun days. We have further discussed the motives behind our parameter values in Appendix \ref{sec:appA}.}

\begin{figure*}
    \centering
    \includegraphics[width=\linewidth]{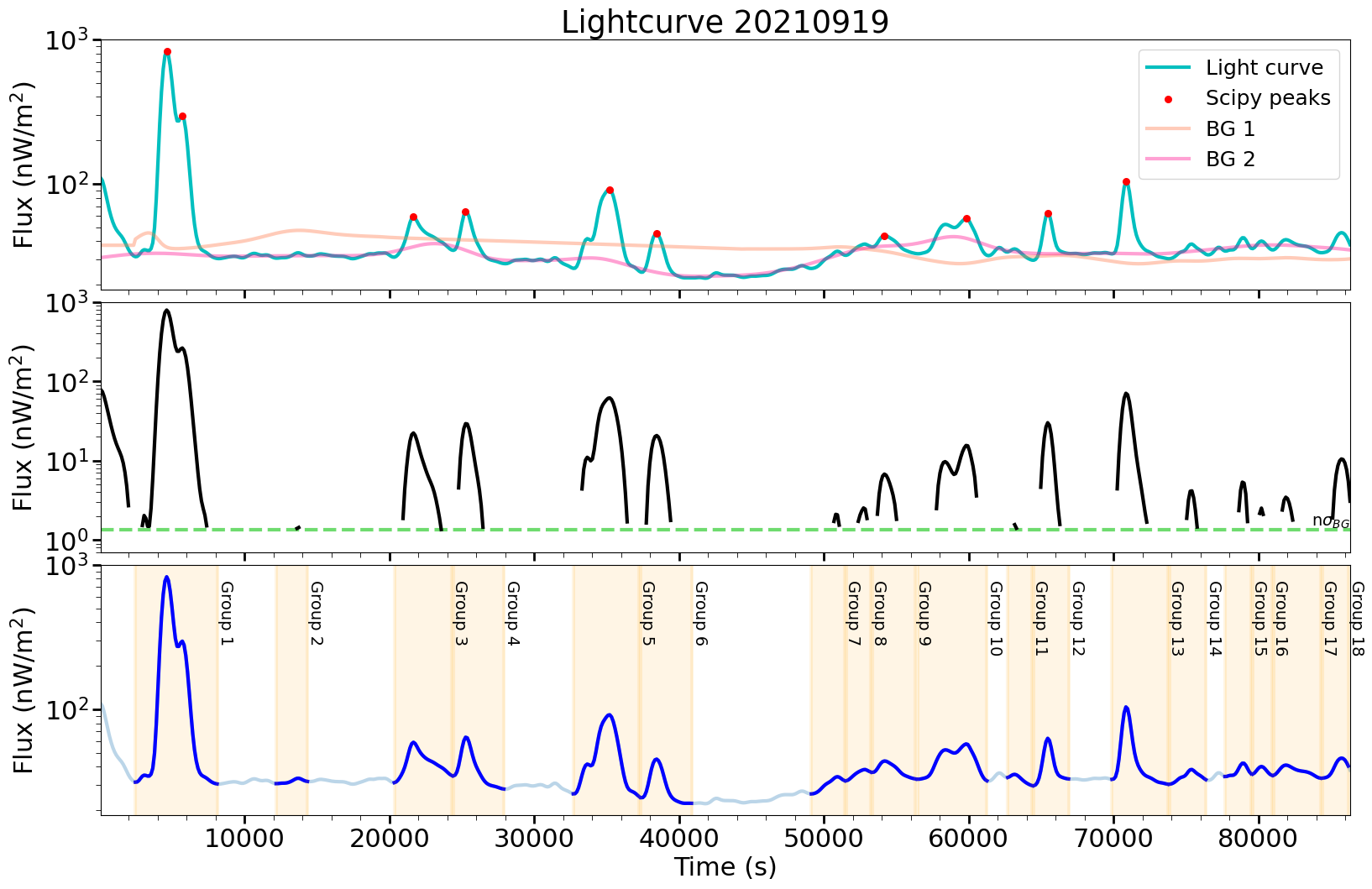}
    \caption{{Top:} Binned and smoothened input light curve 20210919 plotted, in green, in log scale with \texttt{scpeaks} marked in red. The first and second iteration of background estimation are represented by the orange and pink lines respectively. {Middle:} Background subtracted version of the light curve shown above. {The parts of the light curve that do not exceed the dashed green $n\sigma$ line ($n=0.3$) have been removed while the black curves show staggered parts of the light curve above the $n\sigma$ line. Note that as the y-axes of the plots are in log scale, there is an artificial, visual scaling after background subtraction. }
    {Bottom:} Parts of the light curve in light blue show durations without flaring activity while dark blue with yellow highlight shows flare groups. Identified flare groups are numbered chronologically.}
    \label{fig:20210919}
\end{figure*}

We `walk down' the point or group of points exceeding $\tau_{FG}$, independently along both sides, until the count rate reaches the minimum value 0, indicating the end of a flare, or when the slope starts rising again, indicating potential overlap with the next group. To ensure that a `walk down' can always be performed, we linearly interpolate all the data-taking gaps in the data, that are present as \texttt{NaNs}, using the points at the end of these gaps. This is an acceptable fix if the gaps are small and non-contiguous, however, it leads to erroneous detections when the gaps are large and contiguous. As dedicated solar monitoring missions, including XRS/GOES, are not affected by long data-taking gaps induced by orbital considerations, we apply this interpolation in this algorithm implementation. These groups of points are indicated in black in Figure \ref{fig:20210919}. As minor blips may still be present in the smoothened, background-subtracted light curve, the algorithm permits $f$ instances of rising time bins when `walking down' the slope. The region encompassed between the two endpoints is tentatively marked as a flare group. 

The above process is repeated for subsequent points exceeding $\tau_{FG}$. However, we include an additional clause to check if two flare groups are separated by time $< d$. If true, we merge the two flare groups into a single, larger flare group. Although this increases the complexity of the multi-flare decomposition algorithm, this ensures that the identified flares are not assigned an incorrect start or end time, as has been the case in \cite{aschwanden2012_37years} and \cite{goodman2019goes}. 

On active days, the algorithm may not adequately estimate the variations in the background intensity to separate flare groups in regions of high activity. This results in large flare groups with durations $\gtrsim$ 14,000 s, about $1/6^{\text{th}}$ of a day. In such `complex' flare groups, we split the group at local minima points to ensure that the flare decomposition step does not have to deal with very long duration groups with potentially $\mathcal O(10)$ of flare events in the light curve.

{Although the value of soft X-ray flux is conventionally reported in W/m$^2$, we have chosen to report it in nW/m$^2$ throughout the paper as the value of one observed XSM photon count is of the same order-of-magnitude as 1 nW/m$^2$ in GOES range flux. As shown in \citet[Figure 2]{mithun2020xsmperf}, the response function of the detector is approximately linear, and thus this order-of-magnitude relation also scales to higher fluxes. }

\subsection{Modelling a Solar Flare}

Studies such as \cite{tsuneta1987impulsive, dennis1988solar} have classified solar flares into three types: A, B and C. Among these, type B flares seem to dominate in number and have been routinely detected by previous algorithms such as \cite{aschwanden2012_37years, gryciuk2017flare}. These type B flares are characterised by a rapid rise followed by a slow, exponential decay, corresponding to the energy injection and dissipation processes respectively. {A mathematical model will allow us to systematically infer the energetics and temporal characteristics of a flare,} thus, we fit single flares with the Elementary Flare Profile (EFP) discussed in \cite{gryciuk2017flare}, which is mathematically equivalent to an Exponentially-modified Gaussian function. We use this function in the form, 
\begin{equation}\label{eq:EMG}
\begin{split}
    f(t; A, \mu, \sigma, \tau) = A' &\exp\left( \frac{1}{2} \left( \frac{\sigma}{\tau} \right)^2 - \frac{t - \mu}{\tau} \right)\\ &\times\text{erfc}\left( \frac{1}{\sqrt{2}} \left( \frac{\sigma}{\tau} - \frac{t - \mu}{\sigma} \right) \right),
\end{split}
\end{equation}
where,
\begin{equation}\label{eq:Aprime}
    A' = \frac{A\sigma}{\tau} \sqrt{\frac{\pi}{2}}.
\end{equation}
Here, $A$ is the amplitude, $\sigma$ is the standard deviation and $\mu$ is the mean of the Gaussian function, and $\tau$ is the decay parameter of the exponential function. 

\cite{davenport2014kepler} and \cite{kashapova2021morphology} discuss a two-phase decay process composed of both heating and cooling processes resulting in a double-exponential tail. However, we have not incorporated this effect in our study as our spectral bandwidth is considerably large for these effects to be visible.

On the other hand, flaring activities and brightening effects that deviate from the impulsive energy injection model for type B flares are categorised as type A or type C flares. A fraction of these flaring activities may be caused by the blending of impulsive flares but they may have an intrinsically different origin. The exact source of these deviations is not apparent from X-ray light curves alone and require various probing techniques, including 2D imaging across the electromagnetic spectrum. In the soft X-ray channels, we find that type A flares are well-approximated by a Gaussian function, which can be asymptotically obtained at high $\sigma/\tau$ values of the EFP function. On the other hand, type C flares are too complex to be neatly modelled using soft X-ray light curves alone and typically require multiwavelength observations. 

Using the flare curve fit, start, peak and end times are obtained. The peak time is defined as the time when the curve fit attains its peak. The start and end times are taken at the Full-Width Tenth-Maximum (FWTM) height with respect to the peak of the curve fit after subtracting the background level. This is in contrast with the definitions used in \cite{aschwanden2012_37years} which defines the start time as the instance during flare rise when count is 40\% of the peak flux and end time as 50\% of peak flux during flare decay. On the other hand, \cite{gryciuk2017flare} defines the start and end of a flare based on a $1\sigma$ deviation from the background level.

We calculate the signal-to-noise ratio (SNR) of each flare as defined in \cite{babu1996astrostatistics,bradt2004astronomy},
\begin{equation}\label{eq:SNR}
	\text{SNR} = \sum_{flare\ start}^{flare\ end} S / \sqrt{S + 2B},
\end{equation}
where $S$ stands for the signal and $B$ stands for the background, estimated through curve fitting, at each data point.

\begin{figure}
    \centering
    \includegraphics[width=0.6\linewidth]{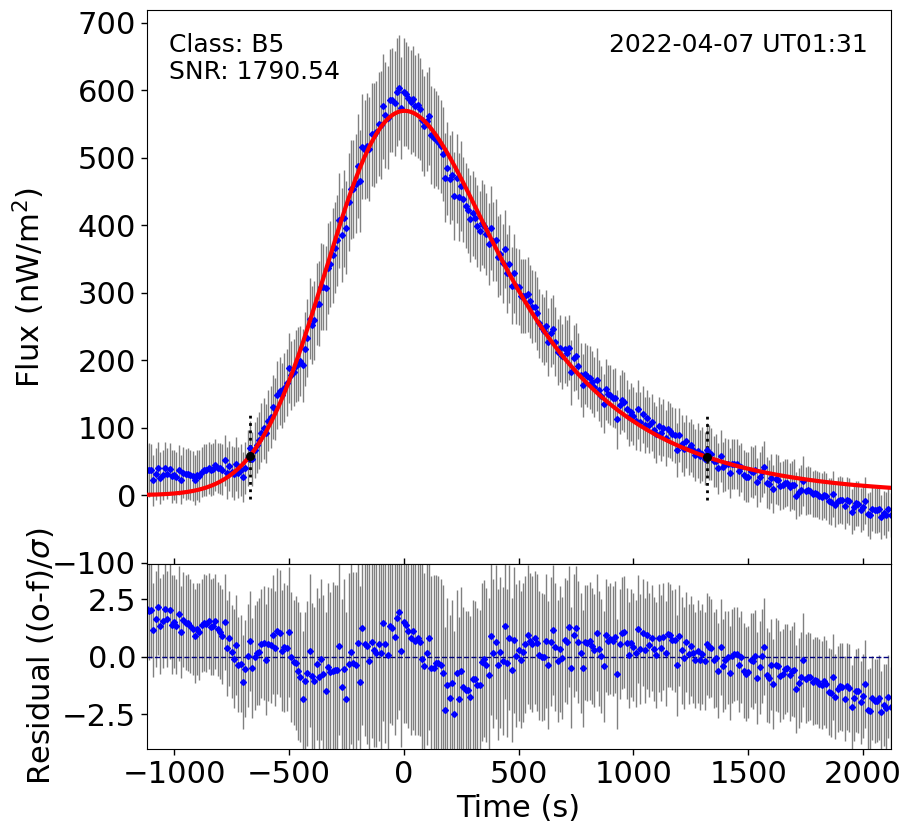}
    \caption{A flare identified by our pipeline. The background-subtracted data points along with their observation errors are marked in blue. The solid red line represents the flare fit, and the dashed vertical black lines indicate the start and end time of the flare at FWTM. The flare peak time is listed on the top right, while the background-subtracted peak flux obtained from the flare fit is used to calculate the flare class listed on the top left along with the SNR calculated using Equation \ref{eq:SNR}.}
    \label{fig:20220407flare}
\end{figure}

\subsection{Decomposition of Multi-Flare Groups}

In cases when it is not possible to fit a flare profile with a single EFP function, we decompose the flare group into multiple, individual flares. The multi-flare is assumed to be a linear superposition of several single flares. However, this may not be true in the case when multiple, overlapping flares are produced by the same flaring location in the Sun, as pointed out in Figure \ref{fig:aia171}, and further research is needed in this topic. 

\begin{figure}
    \centering
    \includegraphics[width=0.6\linewidth]{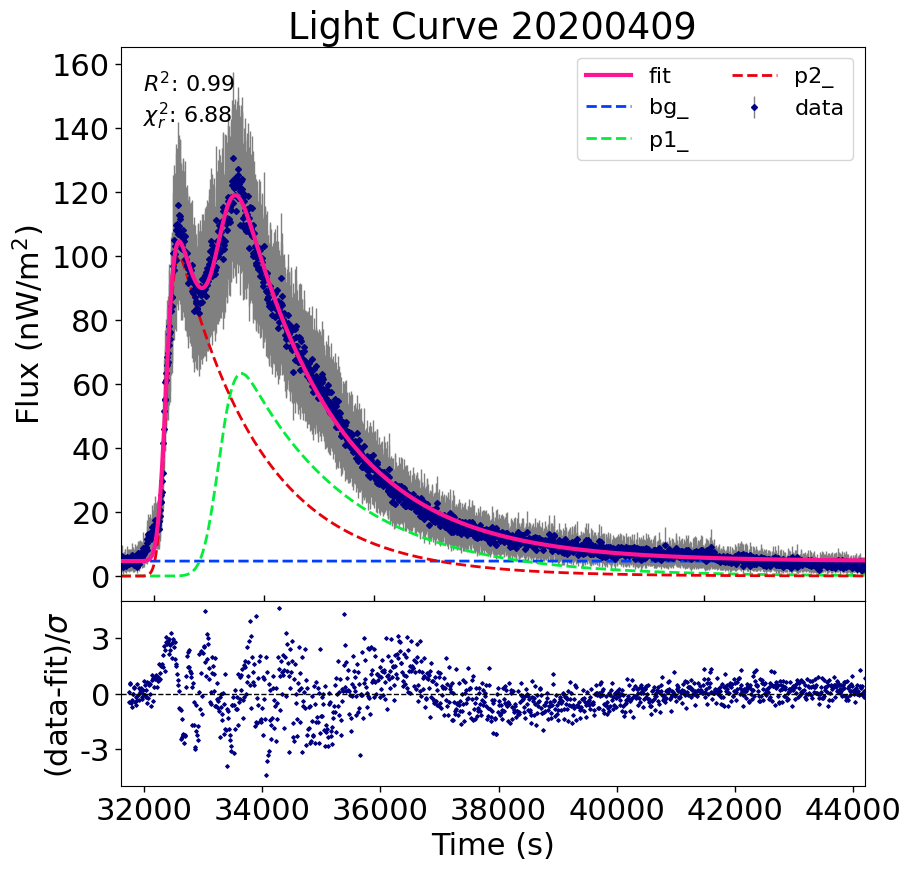}
    \caption{A multi-flare flare group identified in light curve 20200409 decomposed into single flares. 
    {Top:} The observed data points are marked in purple along with their associated observation errors. The dashed lines represent the different components of the fit, with blue representing the background, and red and green depicting the two EFP model fits. The overall fit is given by the solid pink line. The r-squared and reduced chi-square of the obtained fit are stated on the top left. 
    {Bottom:} The residual of the fitted model.
    It can be observed that the green flare p1\_ is actually smaller than the red flare p2\_. }
    \label{fig:20200409reg2}
\end{figure}

\begin{figure*}
    \centering
    \includegraphics[width=\linewidth]{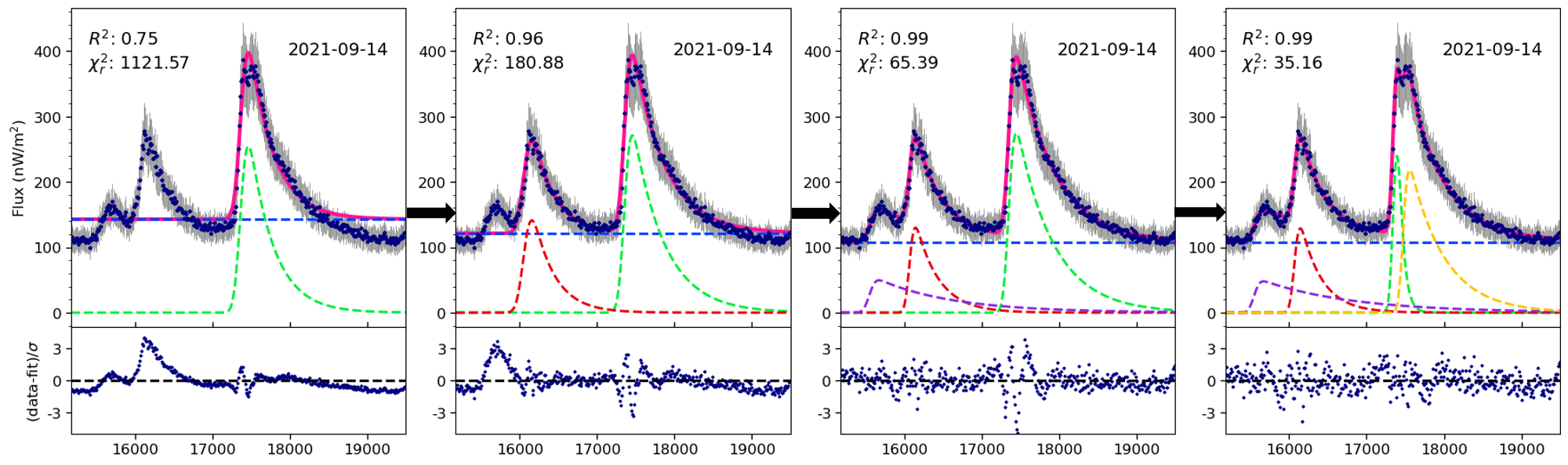}
    \caption{An example of an iterative peak addition process. Each plot follows the same template as Figure \ref{fig:20200409reg2}. On every iteration the reduced $\chi^2$ and R$^2$ values are computed, and once these values meet the threshold criteria or if four peaks have been added, the process is terminated.}
    \label{fig:iterative}
\end{figure*}

We revert to the non-binned, non-smoothened, non-interpolated light curves and use the start and end time estimates of the identified flare groups. For a given flare group, we first estimate the number of single flares needed to describe the flare group. We use the \texttt{scpeaks} module to list peaks and then perform an iterative search over those peaks that exceed a prominence threshold of $0.05$ in normalised flux (an ad-hoc threshold value). Under the assumption that EFP correctly models solar flares, we iteratively add EFP functions, using the python package \texttt{LMfit} \citep{newville13lmfit}, until the chi-square threshold $\tau_{\chi^2}$ and R-squared threshold $\tau_{R^2}$ are satisfied or the residual falls below the minimum error associated with the flux measurement. We also restrict the maximum number of flares within a flare group to four due to limitations on {curve fitting with `too many' free parameters; as EFP is a four-parameter function, attempting to fit a fifth flare is comparable to fitting the data with a $5\times4=20$-parameter equation}. An example of this iterative process is shown in Figure \ref{fig:iterative}. 
As the background of the current flare group may be affected by its occurence within an active region of the Sun, the exponential decay of a previous, larger flare group or a rise in plasma temperature, we model the background as a linear function.

\subsection{Flare Catalog Filtering} \label{sec:methodsfiltering}
The algorithm described in this section is used to generate a candidate list of flares present in the data period. Based on a statistical analysis, we identified three filters to remove erroneous flare detections.  
\begin{itemize} 
    \item Reliability filter: An SNR value below the threshold $\tau_{SNR}=8$, similar to the threshold used in various domains of astrophysics from radio pulsars to gravitational waves are discarded as these are deemed to be unreliable detections. We also filter out flares with $R^2$-value below the threshold $\tau_{R^2_f}=0.5$ \footnote{$\tau_{R^2_f}$ differs from the flare group decomposition threshold $\tau_{R^2}$ as the former is for an individual flare fitted with an EFP function while the latter is for the cumulative fit to a flare group.}. 
    \item Glitch filter: Outlier points in the time series data result in erroneous curve fitting of non-flares. They result in sharply rising, spike-like components, and could be avoided by outlier detection or smoothening the data before curve fitting. However, smoothening will impact curve fitting of small flares as it removes high-frequency components, thus having a negative effect on flare identification. Instead, we discard EFP-fits with $A/\text{peakflux}_{\text{prefit}} < 2$, where $A$ is the amplitude of the flare obtained from EFP curve fitting and $\text{peakflux}_{\text{prefit}}$ is the highest observed flux value within the duration of the flare. 
    \item Background level filter: Due to a rise in plasma temperature or arrival of an active region within the instrument's field of view, the apparent background level can increase. These can also be induced artificially by data-taking gaps, and the movement of the Beryllium filter, employed with a mechanism to limit the flux for higher class flares, in the case of XSM/CH2, and sudden flux drops in the case of XRS/GOES-17; leading to erroneous flares being detected. A few examples are shown in Appendix \ref{sec:appC}. These detections tend to be sigmoid function-like curves and we discard them by setting a threshold on the decay parameter of the curve fit, $\tau_\tau = 10^7$. 
\end{itemize}
The values of these thresholds have been set by inspecting the correlations between various flare parameters. This is elaborated further in Appendix \ref{sec:appC}. 

\subsection{Efficiency of Flare Detection}

The flares detected in XRS/GOES-17 X-ray light curves using the algorithm described in \cite{goodman2019goes} (hereafter referred to as G19) are compared against the flares detected by this algorithm. We also use the catalogs published in \cite{vadawale2021quietsun} (hereafter, V21) and \cite{mondal2021evolution} (hereafter, M21) which specifically contain sub-A and low B class flares (resp.) detected in XSM/CH2 by visual inspection. We call the ratio between the number of detections in these catalogs also being present in our catalog as the `matching rate'. Furthermore, we visually inspect the flares in G19 that are not identified by our algorithm. Such an inspection allows us to optimise the algorithm parameter values, evaluate the performance of the different pipelines, understand the limitations of our algorithm as well as report new detections. We elaborate on these findings in Section \ref{sec:flarevalidation}. 

We also compare our catalog against 2D images from the extreme ultraviolet instrument, Atmospheric Imaging Assembly (AIA), in the Solar Dynamics Observatory (SDO) to determine the validity of a flare detection and estimate the rate of false positives. We inspect a random subsample of A class flares and all sub-A class flares, and present our false positive estimate in Section \ref{sec:falsepositives}.

\section{Results} \label{sec:results}
We generate two flare catalogs using the described algorithm - the first using XRS/GOES-17 data in the period 2020-01-01 to 2022-11-04 (hereafter, SD-GOES) and the second using XSM/CH2 data in the period 2019-09-12 to 2022-11-04 (hereafter, SD-CH2). The catalogs can be found on the code repository accompanying this paper\footnote{\url{https://github.com/DEVANSH-DVJ/SuryaDrishti}. In the main text, we abbreviate \textit{SuryaDrishti} as SD.}. The algorithm settings that were used are listed in Table \ref{tab:consetting} and the catalogs are summarised in Table \ref{tab:catalog}. 

We refer to successive dates without any active regions within the instrument's field of view as Quiet Sun period. We present our flare validation results followed by statistical analyses of different flare classes and types, and peak flux distribution. 

\begin{table}
    \centering
    \begin{tabular}{l c l}
    \hline
    \textbf{Background estimation} & \\
    Binning size & & 12 $\times$ 10 s \\
    Gaussian kernel size & $\sigma_G$ & 2 \\
    Topographic prominence & $\tau_P$     & [4,8,12,23]\tabnote{based on DD season (updated every alternate season to account for variations in solar\\ activity in the course of a solar cycle.)} \\
    Slope & $\tau_m$     & 0.01 \\
    \hline
    \textbf{Flare group identification} & \\
    Modified sigma-clip & $n$ & 0.3 \\
    \multirow{2}{16em}{Minimum separation of neighbouring flare groups} & \multirow{2}{*}{$d$} & \multirow{2}{*}{120 s} \\
     & & \\
    Minimum duration of flare group & & 120 s \\
    Permitted upturns & $f$ & 2 \\
    \hline
    \textbf{Flare decomposition} & \\
    R-squared & $\tau_{R^2}$ & 0.8 \\
    Reduced chisquare & $\chi^2$ & 30 \\
    \hline
    \end{tabular}
    \caption{List of algorithm parameter values used to generate SD-CH2 and SD-GOES.}
    \label{tab:consetting}
\end{table}

\begin{table}
    \centering
    \begin{tabular}{lcc}
        \hline
        Instrument & XSM/CH2 & XRS/GOES-17 \\
        \hline
        Catalog & SD-CH2 & SD-GOES \\
        {Period} & 2019-09-12 to & 2020-01-01 to \\
         & 2022-11-04 & 2022-11-04 \\
        Total Observation Days & 569 & 1038 \\
        Flares Detected & 6266 & 13356 \\
        \hline
    \end{tabular}
    \caption{Summary of the catalogs generated using the pipeline described in this paper.}
    \label{tab:catalog}
\end{table}

\subsection{Flare Validation}\label{sec:flarevalidation}

We validate the accuracy and rate of false positive detections by comparing flares in SD-CH2 and SD-GOES against the flares present in G19 and 94 \AA\ AIA/SDO images. \cite{ryan2012thermal} showed that a previous version of the G19 catalog, with events identified between 1991 and 2007, contained false positives as well as true negatives, setting a lower limit for the latter at 5\%. Thus, we visually inspect a random subsample and present an estimate for the false positive detection rate for our catalog. As a robust understanding of flare number distribution is necessary to obtain a true negatives estimate, we defer such an analysis for a future study.

\subsubsection{Algorithm sensitivity}

\begin{table}
    \centering
    \begin{tabular}{rccccccc}
        \hline
         \multirow{2}{2.5em}{Flare Class} & & \multicolumn{2}{c}{SD-GOES} & \multirow{2}{4em}{\centering SD Not Detected} & \multirow{2}{4.6em}{\centering SD New Detections} & \multirow{2}{4.1em}{\centering Matching Rate} \\
         {} & G19\tabnote{Flares occuring in data-taking gaps have been excluded from this analysis. G19 has\\ 5151 flares in the period 2020-01-01 to 2022-11-04.} & Analysed & Excluded\tabnote{Flares occuring in 2022-08-31 to 2022-10-16 are excluded from this analysis.} & & & \\
        \hline
        Sub-A & 0 & 12 & 0 & 0 & 12\tabnote{Includes false positives as discussed in Section \ref{sec:falsepositives}.} & --\\
        A & 22 & 1850 & 1 & 12 & 1840\tabnote{Includes false positives as discussed in Section \ref{sec:falsepositives}.} & 45.45\% \\
        $\geq$B & 5129 & 10420 & 932 & 483 & 6240 & {90.58\%} \\
        $\geq$C & 2419 & 3201 & 829 & 133 & 1362 & 93.26\% \\
        $\geq$M & 147 & 151 & 31 & 1 & 5 & 99.32\% \\
        X & 8 & 8 & 1 & 0 & 0 & 100\% \\
        \hline
        Total & {5151} & 12282 & 933 & 495 & 8092 & {89.43\%} \\
        \hline
    \end{tabular}
    \caption{Comparison of algorithm sensitivities between G19 and SD-GOES. }
    \label{tab:algosens}
\end{table}

We first present a comparison of flare detections in the SD-GOES and G19 catalogs. This allows us to directly compare the sensitivity of our new pipeline with the GOES algorithm. We queried the G19 catalog\footnote{XRS/GOES data: \url{https://www.ngdc.noaa.gov/stp/satellite/goes-r.html}} using \texttt{SunPy v4.1.0} \citep{sunpy_community2020}, and verified if the peak time of a flare in G19 lies between the start and end times of a flare in SD-GOES. We limit our comparison to days when one-second XRS/GOES-17 light curves are available on the NOAA website, and we exclude flares detected in the period 2022-08-31 to 2022-10-16 as the light curves contained significant data-taking gaps. The results have been summarised in Table \ref{tab:algosens}. Column 5 refers to the flares in G19 not detected by our pipeline with the corresponding matching rate in Column 7, assuming none of them are false positives. Column 6 lists the number of flares in SD-GOES that are not present in G19.

We identify 89.64\% greater than B1.0 class, including all X class and all but one M class flares. Among the 495 flares that we could not validate, 26 flares (all under B class) were filtered out in the flare catalog filtering step. This is much smaller than the 6240 new, greater than B1.0 class flares we report that are not present in the G19 catalog. As shown in Figure \ref{fig:fourregions}, we find many similar instances where groups of flares were not decomposed into single flares resulting in an overestimation of the population of large flares in G19. 

Among A class flares, although we only detect 10 / 22 of the A class flares present in G19, we infer from visual inspection that 6 / 12 non-detections are in fact false positives. Thus, we deduce a matching rate of 62.50\% (10 / 16). Once again, we report many more new A and sub-A class flares in SD-GOES, that are not present in G19. In the following subsection, we verify the correctness of these new detections.

\subsubsection{Rate of false positives}\label{sec:falsepositives}

We consider all sub-A class flares and a random subsample of 50 A class flares each from SD-CH2 and SD-GOES, and compare them against 94 \AA\ images from AIA/SDO. We manually accessed 94 \AA\ images\footnote{AIA/SDO data: \url{http://jsoc.stanford.edu/data/aia/images/}}, and employed the AIA flare localisation method described in \cite{vadawale2021quietsun} for flare validation: we scroll through consecutive images and infer a flare if a sudden brightening or curling line brighter than the solar background occurs at $\pm$ 120 seconds. If no such region is visually identified, we assume no flare took place within that time duration. This is used to estimate the rate of false positive detections {for weaker flares.} 

Among sub-A class flare detections, we find that 3 / 12 (25\%) in SD-GOES and 10 / 213 (4.7\%) are false positives. As XRS/GOES-17 is not rated to be sensitive to changes $< 10^{-8}$ W/m$^2$, a high rate of false positives were expected in its sub-A class regime. 

On the other hand, in our random subsample of 50 A class flares, we could not find a corresponding brightening in the AIA/SDO images in 11 / 50 (22\%) SD-GOES subsample and 1 / 50 (2\%) SD-CH2 subsample. Most of these false positive detections follow a characteristic long decay phase, triggered by an increase in the background intensity. This is similar to, but not as extreme as, the erroneous flare detections discussed in Figure \ref{fig:tauexamples} of Appendix \ref{sec:appC}, and might be removable with more precise catalog filters. 

{Given the $\gtrsim 90\%$ matching rate of SD-GOES with the G19 catalog, we do not carry out this analysis for B class and stronger flares, and assume that they only contain a small number ($< 1\%$) of false positives. Thus, we estimate an overall false positive rate of $1.33\%$ for the entire SD-CH2 catalog, by accounting for the 10 sub-A class false positives, a 2\% false positive rate for A class and a conservative upper limit of 1\% false positive rate for B class and stronger flares.}

\subsubsection{Comparison with other flare catalogs}

Apart from flares in G19 and AIA/SDO images, we compared our catalog with some recent works that used automatic or manual detection methods. \cite{plutino2023new} claims 5$\times$ more detections on GOES data over a 34 year period, however, a large set of continuous detections between 2019-09-12 to 2019-10-12, which overlaps with our observation period, had an apparent start time of 9 AM. We believe there is a systemic error in the detection or classification system used by these authors, at least during the Quiet Sun days. Therefore, at the time of writing, we have not compared their results with our catalog. 

The V21 catalog has 98 visually identified microflares\footnote{{Sub-A class flares are sometimes referred to as \textit{microflares} in the literature.}} between 2019-09-12 and 2020-04-30, corresponding to the first two D-D seasons of XSM observations, which overlaps with the 2019--2020 solar minimum. When we compared SD-CH2 against V21, we had detected 11 of the 21 $>$A0.2 flares. We could not automatically detect any flare weaker than A0.2, which correspond to less than 1 photon count per second. Microflares are more prominent at lower energy radiation (EUV and 1--1.5 keV) than higher energy radiation. The inability to automatically detect $<$A0.2 flares could be a direct consequence of the smaller energy bandwidth 1.5--12.4 keV used in this work compared to 1--15 keV bandwidth used in V21.  When we compared SD-CH2 against the M21 catalog which contains nine low B class flares from the same Quiet Sun period as above, SD-CH2 contained all nine of them. 

\subsection{Flare Catalog}

\begin{table*}
    \centering
    \begin{tabular}{rcccccc}
        \hline
         {Flare} & \multicolumn{3}{c}{SD-CH2} & \multicolumn{3}{c}{SD-GOES} \\
         {Class} & Type A & Type B & Total & Type A & Type B & Total \\
         \hline
        Sub-A & 41 & 172 & 213 & 2 & 10 & 12 \\
        A & 318 & 1012 & 1330 & 559 & 1292 & 1851 \\
        $\geq$B & 1110 & 3613 & 4723 & 2478 & 8874 & 11352 \\
        $\geq$C & 203 & 1038 & 1241 & 699 & 3331 & 4030 \\
        $\geq$M & 3 & 66 & 69 & 7 & 175 & 182 \\
        X & 0 & 0 & 0\tabnote{The preprocessed light curves with X class flares were not available on the PRADAN website\\ at the time of analysis.} & 0 & 9 & 9\\
        \hline
        Total & 1469 & 4797 & 6266 & 3039 & 10176 & 13215 \\
        \hline
    \end{tabular}
    \caption{Breakdown of the SD-CH2 and SD-GOES flare catalogs into flare classes and flare types.}
    \label{tab:flarecat}
\end{table*}

\begin{figure*}
    \centering
    \includegraphics[width=\linewidth]{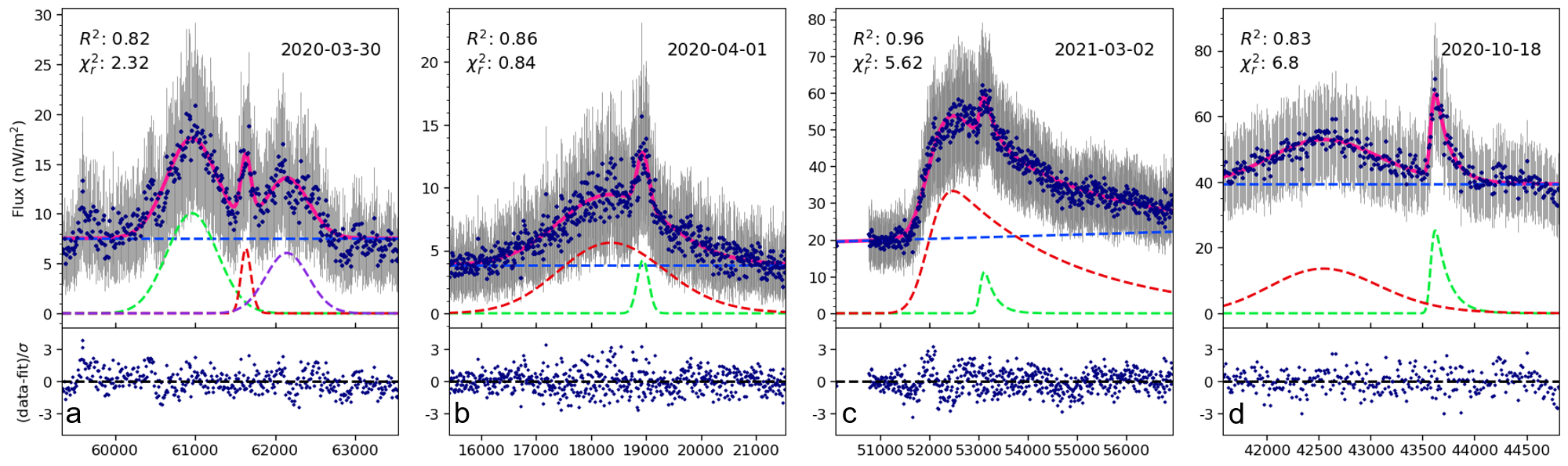}
    \includegraphics[width=\linewidth]{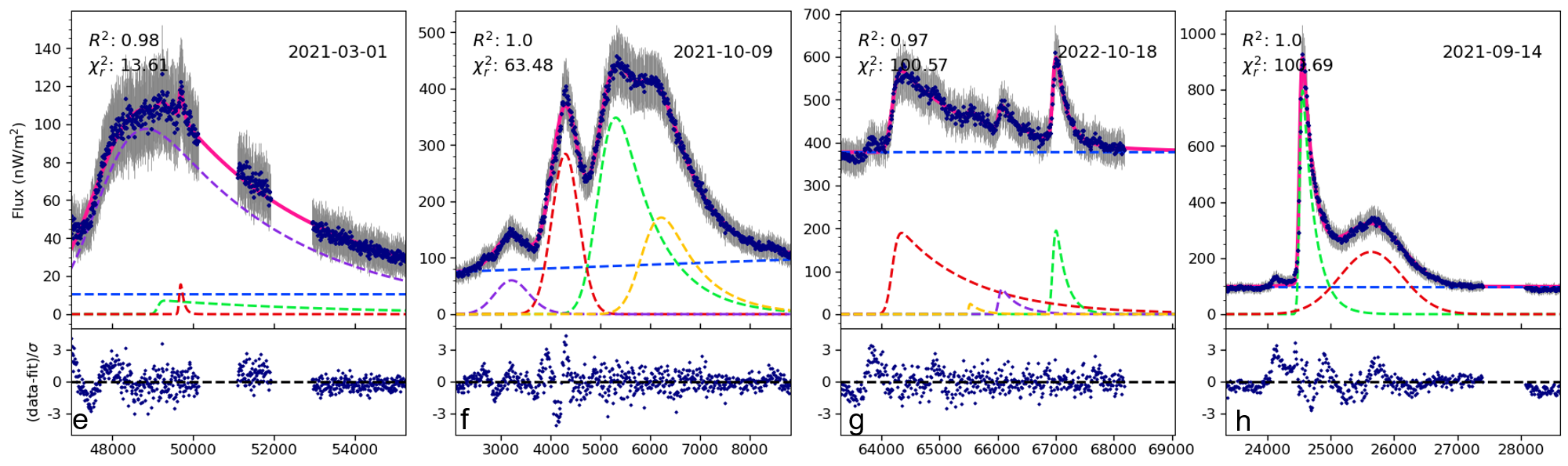}
    \includegraphics[width=\linewidth]{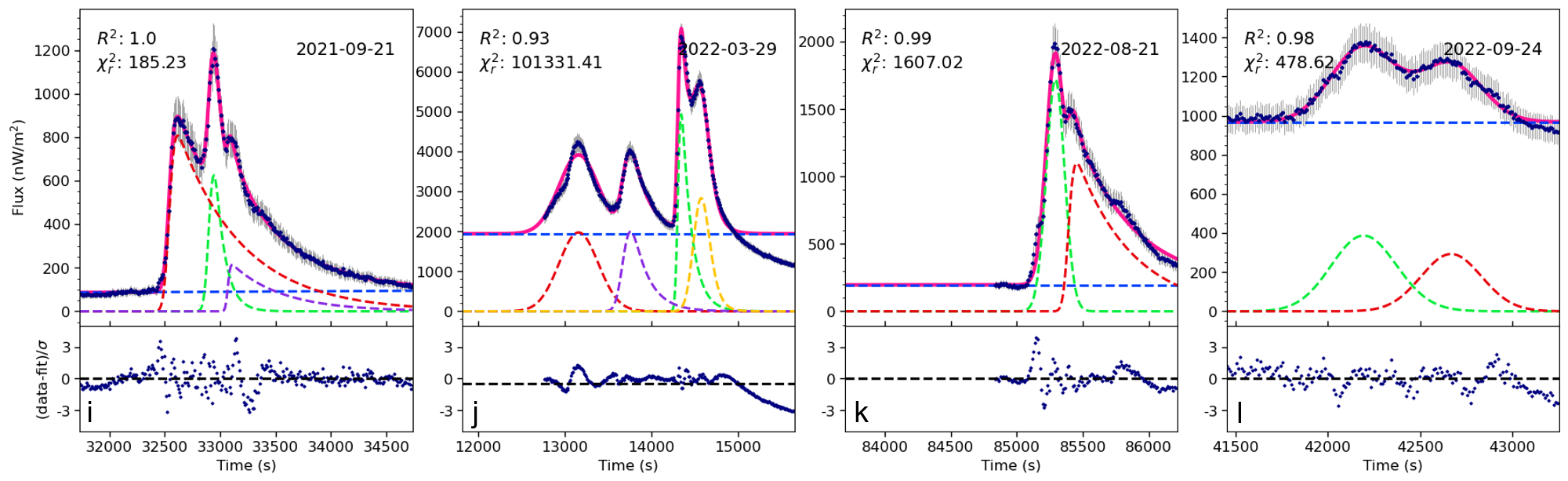}
    \caption{Examples of multi-flare groups decomposed into individual flares from SD-CH2. Each plot follows the same template as Figure \ref{fig:20200409reg2} with the dashed lines representing the different components of the fit. The largest flare in each row from the top is of class A, B and C respectively and the rightmost column (plots d,h,l) in each row contains at least one type A flare. }
    \label{fig:fourregions}
\end{figure*}

\begin{figure*}
    \centering
    \includegraphics[width=\linewidth]{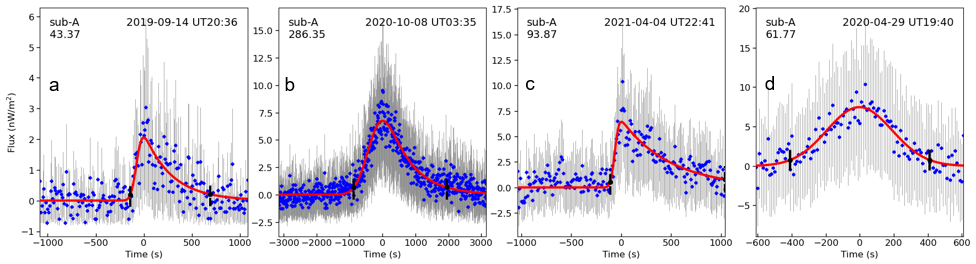}
    \includegraphics[width=\linewidth]{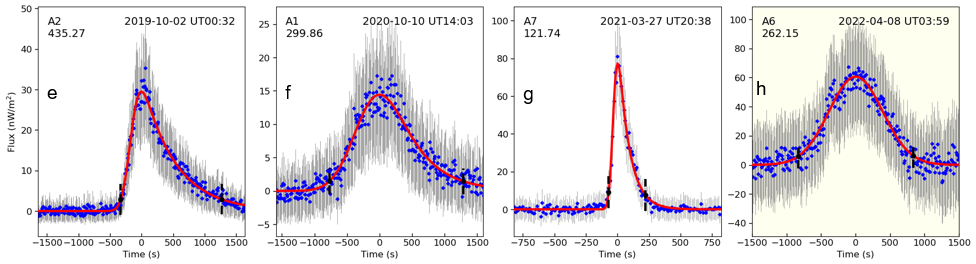}
    \includegraphics[width=\linewidth]{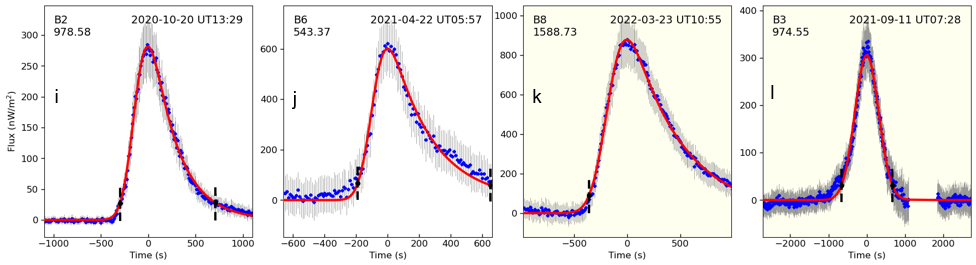}
    \includegraphics[width=\linewidth]{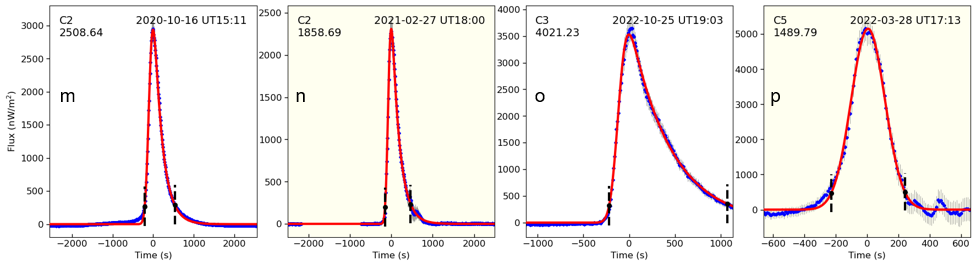}
    \caption{Examples of single flares along with their EFP model fits. Each plot follows the same template as Figure \ref{fig:20220407flare}. The tinted plots depict flares which were decomposed from different multi-flare groups. From the top, each row present class sub-A, A, B and C respectively and the rightmost column (plots d,h,l,p) shows type A flares. }
    \label{fig:fourflares}
\end{figure*}

We present the contents of the SD-CH2 catalog which includes temporal, energetics and background characteristics of 6266 solar flares derived from their EFP model fits. During the period considered for SD-CH2, 569 days exceeded the necessary 40\% temporal coverage. The catalog contains entries for each flare, with multi-flare groups being decomposed into individual flares. Figure \ref{fig:fourregions} shows an array of decomposed multi-flare groups and Figure \ref{fig:fourflares} shows an array of background-subtracted single flares, including those obtained from the decomposition of multi-flare groups. 

The contents of the catalog have been summarised in Table \ref{tab:flarecat}. We also provide a summary of the contents of SD-GOES to facilitate comparisons. An excerpt from the full catalog can be found under the \textit{supplementary material} accompanying this paper. We show all fields for four catalog entries: the first entry is a single flare, the second is of a decomposed flare from a multi-flare group and the remaining two are the decomposed flares from the same multi-flare group. The entire catalog can be accessed on our code repository. 

\subsubsection{Automatic detection of A and sub-A class flares}

SD-CH2 contains 1330 A and 213 sub-A class flares in three years of observations, which includes large swaths of quiet days. In the first 365 days of observations, which was close to the solar minima, we detected 153 A and 156 sub-A class flares. The first row (plots a--d) in Figure \ref{fig:fourregions} contains examples of decomposed multi-flare groups where the largest flare is smaller than B1.0, and, the first two rows (plots a--h) in Figure \ref{fig:fourflares} contains examples of background-subtracted single sub-A and A class flares. 

\subsubsection{Bimodality in the rate of flare rise}

\begin{figure}
    \centering
    \includegraphics[width=0.49\linewidth]{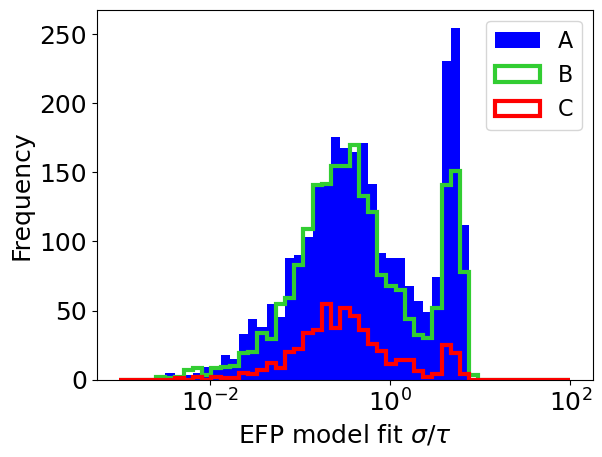}
    \includegraphics[width=0.49\linewidth]{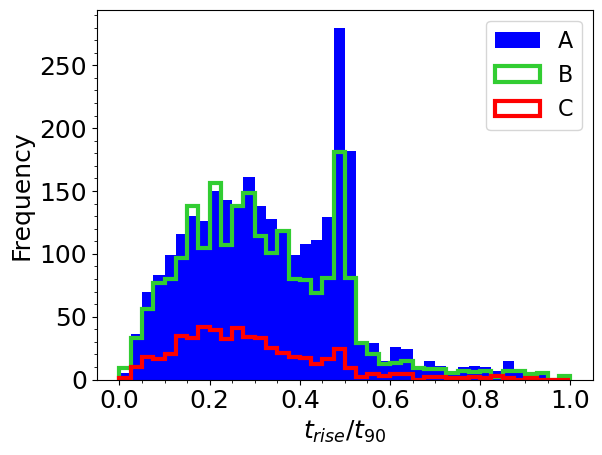}
    \caption{Temporal characteristics of flares with different colours corresponding to different flare classes ($<$A and $>$C classes are not plotted). {Left: }Ratio of EFP function temporal parameters $\sigma$ and $\tau$ with a valley at $\sigma/\tau \sim 2$. {Right: }Fraction of flare duration in rising phase.}
    \label{fig:sigmatau}
\end{figure}

\cite{tanaka1983types, tanaka1987impact, tsuneta1987impulsive, dennis1988solar} were one of the earliest to point out the existence of different types of solar flares, with most flares having a fast rise followed by an exponential decay while some have a slow rise and are more Gaussian-like. We echo similar findings and present the first large-scale catalog of slow rise flares. We see a clear bimodality in the $\sigma/\tau$ distribution, shown on the left in Figure \ref{fig:sigmatau} which corresponds to a dimensionless ratio between the width and decay rate of flares, and serves as a proxy to determine the rate of rise of flares. This bimodality is independent of flare class and has a cutoff at $\sigma/\tau \sim 2$. Flares with $\sigma/\tau \lesssim 2$ correspond to the impulsive or type B flares that have been routinely identified and widely studied. Similar inferences can be drawn from the plot on the right in Figure \ref{fig:sigmatau} which shows the fraction of time spent by a flare in the rising phase. There is a clear peak at $\sim 0.5$ which indicates Gaussian-like flares. The rightmost column in Figures \ref{fig:fourregions} and \ref{fig:fourflares} illustrate such gradual-rise flares, also classified as type A, and we discover a total of 1469 such flares.

The flare detection algorithms described by \cite{aschwanden2012_37years} and \cite{goodman2019goes} exploit the knowledge of rate of flare rise to detect type B solar flares. They look for an exponential rise with an index of at least 1.4 and 1.225 respectively. However, type A flares do not have a steep rise and cannot be easily detected by these algorithms. Thus, our work provides a useful platform to select and study these type A flares. The above algorithms also expect that consecutive flares are separated well-enough such that the rising phase and flux peak are distinct. This is not the case in flare groups with shoulder peaks which our algorithm would decompose into individual flares.

\cite{tsuneta1987impulsive} had also pointed out a third class of flares, referred to as type C flares, which do not strictly fall under type A or B. Our algorithm identifies some `complex' flare groups, which may correspond to these type C flares, and proceeds to decompose these groups into individual flares using the EFP function. They are artificially forced into the type A or type B bucket. The lack of a specific, X-ray flux signature indicates that type C flares are not well modelled by linear superpositions of EFP functions and require further investigation of their integrated light curve properties. 

\subsection{Peak Flux Distribution} 

\begin{figure}
    \centering
    \includegraphics[width=0.49\linewidth]{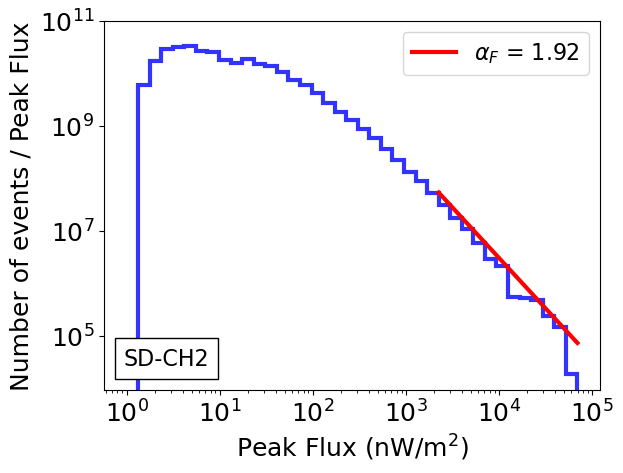}
    \includegraphics[width=0.49\linewidth]{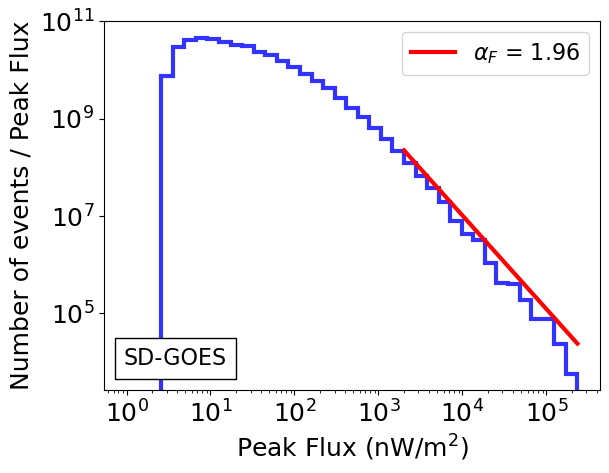}
    \caption{Background-subtracted peak flux distributions with power law fits of flares in SD-CH2 (left) and SD-GOES (right). {The power law fits (in red) have been overlayed on this plot using the value inferred from Appendix \ref{sec:appC}}.}
    \label{fig:powerlaw}
\end{figure}

We analyse the exponent $\alpha$ in a power law fit $Ax^{-\alpha}$ to both with-background and background-subtracted peak flux distributions\footnote{$\alpha$ is sometimes referred to as the power index.}. As is common in the solar physics literature, we have plotted the histogram (i) without cumulative addition -- {to examine the number of peak fluxes within a specific energy bin}, and after (ii) scaling the bin heights by $1/\text{bin width}$ -- {to account for the increasing bin sizes at higher peak fluxes, as the bin widths are equally spaced in log scale.}

\begin{table*}
    \centering
    \begin{tabular}{ccccccc}
        \hline
         {} & \multicolumn{3}{c}{SD-CH2} & \multicolumn{3}{c}{SD-GOES} \\
         $\alpha_F$ & Type A & Type B & All & Type A & Type B & All \\
        \hline
        with-background & 1.89 $\pm$ 0.06 & 2.23 $\pm$ 0.10 & 2.30 $\pm$ 0.08 & --\tabnote{Shows no plateau of stability.} & 2.19 $\pm$ 0.08 & 2.30 $\pm$ 0.10\tabnote{Shows a poor plateau of stability.} \\
        background-subtracted & 1.94 $\pm$ 0.08 & 1.90 $\pm$ 0.09 & 1.92 $\pm$ 0.09 & 1.84 $\pm$ 0.08 & 1.97 $\pm$ 0.07 & 1.96 $\pm$ 0.08 \\
        \hline
    \end{tabular}
    \caption{Power indices of curve fits to with-background peak flux and background-subtracted peak flux.}
    \label{tab:powerlaws}
\end{table*}

As we increased the value of the `turnover' point $x_{min}$ in the curve fit, we found that the value of $\alpha_F$ also varied. This called for a need to analyse the stability of the power-law fit. Thus, we perform a Monte-Carlo analysis to confirm our intermediate outcome. We elaborate on our methods in Appendix \ref{sec:montecarlo}.

From our Monte-Carlo analysis, we infer $\alpha_F = 1.92 \pm 0.09$ and $\alpha_{F} = 1.96 \pm 0.08$ for the background-subtracted peak flux distribution from SD-CH2 and SD-GOES respectively. The turnover point $x_{min} \sim 2\times 10^{-6}$W/m$^2$. We present these histogram distributions along with power law curve fits in Figure \ref{fig:powerlaw}, and summarise the values, following the same Monte-Carlo method, for other supposedly power law distributions in Table \ref{tab:powerlaws}. The implications of these values are discussed in Section \ref{sec:discussionspowerlaw}.

\section{Discussion} \label{sec:discussions}
Our catalog presents 6266 solar flares detected over a 38-month period, of which 1469 are type A and 4797 are type B flares. This catalog also includes 1330 A class flares and 213 microflares making it the most extensive weak flare catalog detected by a single instrument. We have measured flare properties in a systematic, uniform way. As such, the catalog represents a unique and useful resource for studying the solar flare population, such as statistical comparisons between type A and B flares. Future missions like Aditya-L1 will extend this catalog to a larger database with better sensitivity and SNR. 

Additional analyses of the catalog are ongoing, including the determination of temperature and emission measures, as well as studying different flare morphologies. {The $\gtrsim 90\%$ matching rate with the G19 catalog suggests that no new systemic biases have been introduced in the detection of large, type B flares, while improving the ability to detect type A flares. Nevertheless, it is important to address the $\sim 10\%$ discrepancy between G19 and SD-GOES in B class flares and determine possible selection effects induced by our algorithm design and observation method, including the effects of flare decomposition on the apparent flare number distribution. Such an analysis may also reveal new flare morphologies.}

In this section, we discuss the main results of this paper, namely, the performance of the instrument and the algorithm during the solar minima, the implications of a large catalog of microflares and type A flares, the power law fit to the peak flux distribution {and the implications and optimism surrounding future missions with imaging capabilities in the X-ray regime}. 

\subsection{Population of A and Sub-A Class Flares}

Given our algorithm's emphasis on correctness and from our estimate of the rate of false positives, we can be confident about robust detections of A and sub-A class flares during Quiet Sun periods. Currently, over 70\% of the detected sub-A class flares, occurred within the first year of observation, which is considered by many to be one of the deepest solar minimum of the past century. Our results corroborate that XSM/CH2 is a more sensitive instrument than XRS/GOES-17 given its ability to discern smaller increases in flux, resulting in fewer false positive A and sub-A class flare detections as well as a higher daily-rate of such detections. 

{Although we have detected a significant number of sub-A class flares, our catalog indicates a selection effect arising due to the concentration of microflare detections on quiet days. Thus, the true rate of microflares is expected to be much higher.} Instruments with better sensitivity and X-ray imaging capabilities will only advance the knowledge of these class of flares which are rarely resolved with existing instruments outside the Quiet Sun period. 

\subsection{Population of Type A and B Flares}

This work unlocks a new step in studying a distinct type of flares. Spatial and spectroscopic observations of type A flares can be used to further study the evolution of various plasma parameters, its resemblance and correlation with other high-energy solar events, the solar coronal magnetic field, and other statistical properties. 

Flare morphologies different from that of type A and B are not consistently identified by the algorithm. Type C flares, which are detected, are currently forced into one of those two buckets. Furthermore, flaring activities that cannot be modelled by an EFP function are filtered out. This may include some type C flares and possibly other yet-to-be discovered phenomena, and are reserved for future work. In its current form, the algorithm is not designed to detect type C flares and more work needs to be done to ensure they are distinctly identified.

\subsection{Peak Flux Distribution}\label{sec:discussionspowerlaw}

Given the use of a flare decomposition step, as well as higher sensitivities for both our algorithm and our instrument, we expected to detect a higher number of weaker flares than previously discovered thereby leading to an $\alpha_F > 2$. Indeed, the turnover point has moved leftwards and the power index of the with-background peak flux distribution is slightly higher than previously reported. 

On the other hand, with regards to the background-subtracted distributions, we have arrived at a power index which is roughly consistent with the value $\alpha_F \sim 2$. The similarity of power indices for type A and B flares suggest that the underlying generative process may still be the same. \cite{aschwanden2012soc} presents a theoretical model to explain this power index. 

It should be noted that some authors have already called into question the validity of a power law fit \citep{wheatland2010evidence, ryan2016effects,verbeeck2019solar}. We managed to reproduce results similar to \citet[Figure 5]{ryan2016effects} (refer Appendix \ref{sec:montecarlo}), and thus, echo the arguments presented in \citet[\S4.2]{ryan2016effects}. We only find a weak plateau of stability in the range $x_{min} \in [1.4, 3.1] \times 10^{-6}$W/m$^2$, which is left of the turnover point $x_{min} \sim 5\times10^{-6}$W/m$^2$ mentioned in \cite{ryan2016effects}. Although our study uses a different flare detection method, it is still possible that a detection bias is present, especially at $<10^{-6}$W/m$^2$. Analyses of the true number distribution of solar flares with higher sensitivity instruments and a year-wise progression of the peak flux distribution, similar to \citet{aschwanden2012_37years}, will help resolve this conflict.

\subsection{Implications for Imaging Telescopes}

{Improved instrumentation techniques will help us understand the limitations and assess the systemic biases that may be present in the analyses of time-series data. Adding X-ray imaging capabilities to the existing soft X-ray spectroscopy data will greatly enhance many aspects of research in solar flares and will extend the results in \citet{battaglia2012rhessi,petkaki2012sdo} in the extreme ultraviolet regime to higher energy bands: (i) the background can be handled independently and in a more methodical manner; (ii) the region surrounding a flare can be extracted and studied separately; (iii) the regional activity on the corona, from the base of the flare to the top of the coronal loop, and the processes involved in generating different flare intensities and morphologies can be studied; and {(iv) individual active regions can be analysed and temporal distribution of multiple, overlapping flares from different active regions can be explored} provided detailed spatial resolution is available.}

\section{Conclusions} \label{sec:conclusion}
We have presented a large sample of solar flares, and for each entry, we provide detailed temporal and background properties with the help of an EFP-model fit. This yields a large sample of flares, both type A and B, enabling a direct comparison between them. This data set will be explored in further detailed studies by our team. The statistics of flare spatiotemporal morphology, the spatial distribution of flares with respect to the Sun angle, and a study of the selection effects at play in calculating peak flux distributions will be performed. {A robust understanding of the number distribution of solar flares will be especially useful as the power law fit of peak flux distribution is being repeatedly challenged.} 

{The true number distribution will also be useful to validate the sensitivity of our algorithm. There are a few known limitations in the version of the catalog presented in the paper: (i) we have limited the maximum number of decomposed flares within a flare group to four. This may have led to an underestimation in number of flares within long duration flare groups; (ii) most sub-A class detections occured within the first year of observations
, indicating a selection effect favouring quiet days for weak flare detection; (iii) the $\sim 10\%$ discrepancy between G19 and SD-GOES suggests a higher than previously estimated false positive rate in G19 \citep{ryan2016effects} or a yet-to-be ascertained selection effect in our pipeline; and (iv) some possibly type C flares may have been artificially classified as type A or B flares.}

Nevertheless, we have demonstrated a significant step forward in the number and morphology of flare detections while reducing the rate of false positives. We invite and look forward to the solar physics community making use of this first XSM/CH2 catalog for new interpretations of our results. The release of this catalog will be closely followed by more detailed observations using the SoLEXS instrument onboard the Aditya-L1 spacecraft, which will enable more new and exciting science. 

%

%

%
\appendix   
\section{Algorithm parameters and runtime} \label{sec:appA}

A short description of the algorithm parameters is presented here. The relevant sections of the main text explain the parameters in detail and the employed parameter values have been summarised in Table \ref{tab:consetting}. The code took $\sim30$ minutes to run over the XSM dataset when parallelised across six \texttt{AMD\textregistered\ Ryzen 7} CPU cores and has been tested to work on both Windows and Ubuntu operating systems. This translates to $\sim20$ seconds per light curve.

The binning size refers to the number of datapoints that get bin-averaged together, and the Gaussian kernel refers to the size of a normalised Gaussian that is used to smoothen the data used in the background estimation and flare group identification steps. Higher bin size removes high frequency noise but results in lower sensitivity to small changes. $\sigma_G=2$ corresponds to roughly 41 discrete bins to be sampled for smoothening with a higher $\sigma_G$ value weakening the ability to detect small changes. 
Topographic prominence is an argument for the \texttt{scpeaks} module to identify an initial set of peaks in the light curve. It sets the threshold which a peak has to exceed to be identified. The slope threshold works in the same manner as the slope threshold in \cite{aschwanden2012_37years}. We take the ratio of the first and fourth point in a stream of monotonically increasing flux. If they exceed the slope threshold, it indicates the presence of a flaring duration. 

The modified sigma-clip sets the minimum number of standard deviations of the background estimate that needs to be exceeded for a group of data points to be identified as a flare group. A higher value leads to more robust detections but at the cost of failing to detect small peaks. If two flare groups are separated by less than the minimum separation, then they are merged into one group. Setting a high minimum will result in more `complex' flare groups and a longer flare decomposition runtime. If a flare group is smaller than the minimum duration, it is discarded. The smallest detectable flares are in the timescale of a few minutes. Permitted upturns refers to the number of increasing data points allowed when identifying flare groups through a `walk down'. They help in merging shoulder peaks to within one flare group which allows for a systematic calculation of flare's temporal characteristics. The iterative addition of peaks in the flare decomposition step is terminated when the r-squared and reduced chisquare thresholds are met. A higher threshold will result in more non-decomposed flares. However, most large $>$C class flares, will not meet the reduced chisquare threshold even with a visually indistinguishable EFP fit as the calculation is highly nonlinear.

\section{Setting flare filter thresholds}\label{sec:appC}

\begin{figure}
    \centering
    \includegraphics[width=0.6\linewidth]{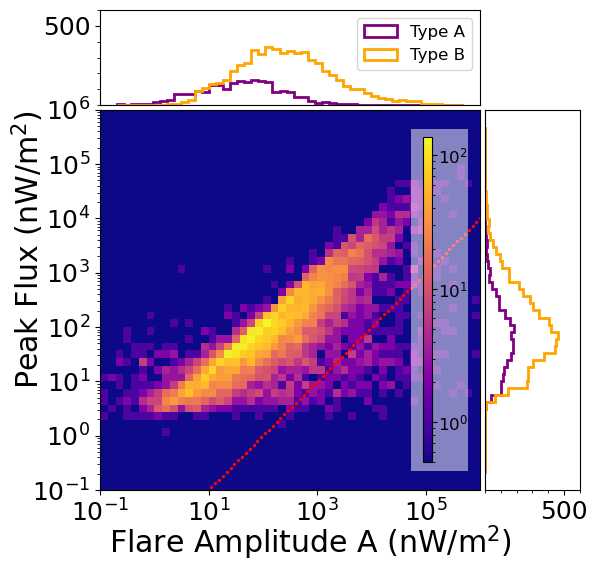}
    \caption{2D histogram of flare amplitude and pre-fit peak flux in log-log scale. The dotted red line splits the plot at A/$\text{peakflux}_{\text{prefit}}$ = 100. }
    \label{fig:Avspeakflux}
\end{figure}

\begin{figure*}
    \centering
    \includegraphics[width=\linewidth]{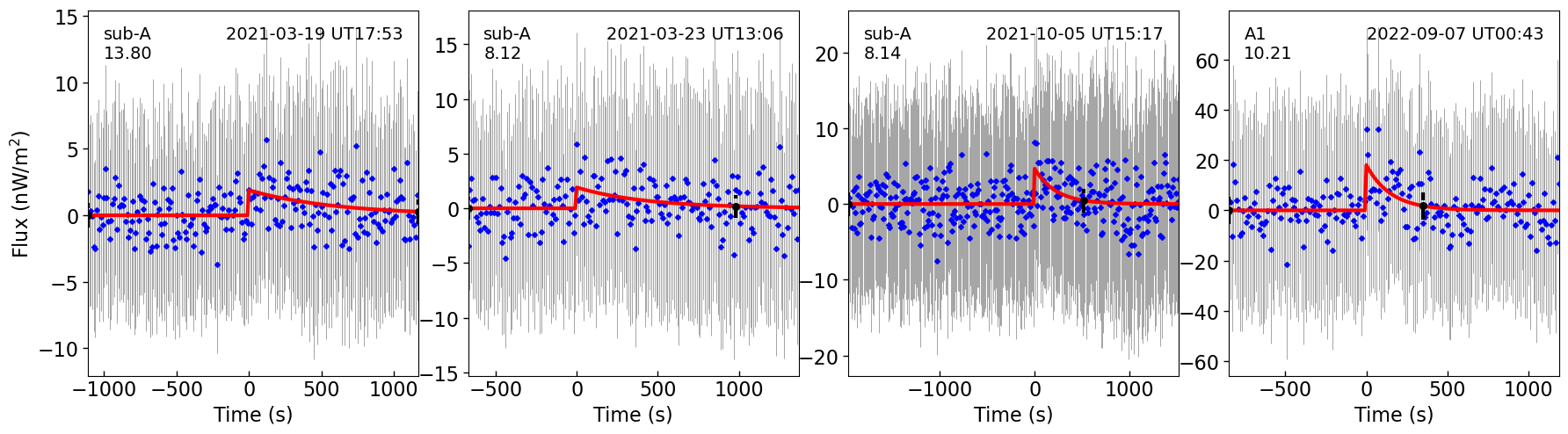}
    \caption{Erroneous flares detected due to glitches in the time-series data in SD-CH2 due to outlier points and constrained flare decomposition.}
    \label{fig:amppeak}
\end{figure*}

We performed a cold run of the pipeline with conservative parameter values across the entire dataset. We discussed the effects of the various algorithm parameters in Appendix \ref{sec:appA}. The result of this run motivated the search for flare catalog filters and these have been applied to both SD-CH2 and SD-GOES.

We compared the relationship between curve fit parameters, identified outliers and inspected their correctness. In Figure \ref{fig:Avspeakflux}, we find outliers to one side of an otherwise linear relationship between the peak flux of a flare and the EFP model-fit amplitude. The effects of the glitch filter are more prominent at lower fluence due to higher variability in peak fluxes for weaker flares. Examples of flares that do not pass this filter are shown in Figure \ref{fig:amppeak}. In many instances, they result from the inflexible EFP fit constraints used in flare decomposition after finding potential peaks in the flare group using \texttt{scpeaks}. This necessitates a flare fit with a peak time close to the peak found by \texttt{scpeaks} even if the time series data cannot be modelled by an exponentially modified Gaussian. 

\begin{figure*}
    \centering
    \includegraphics[width=\linewidth]{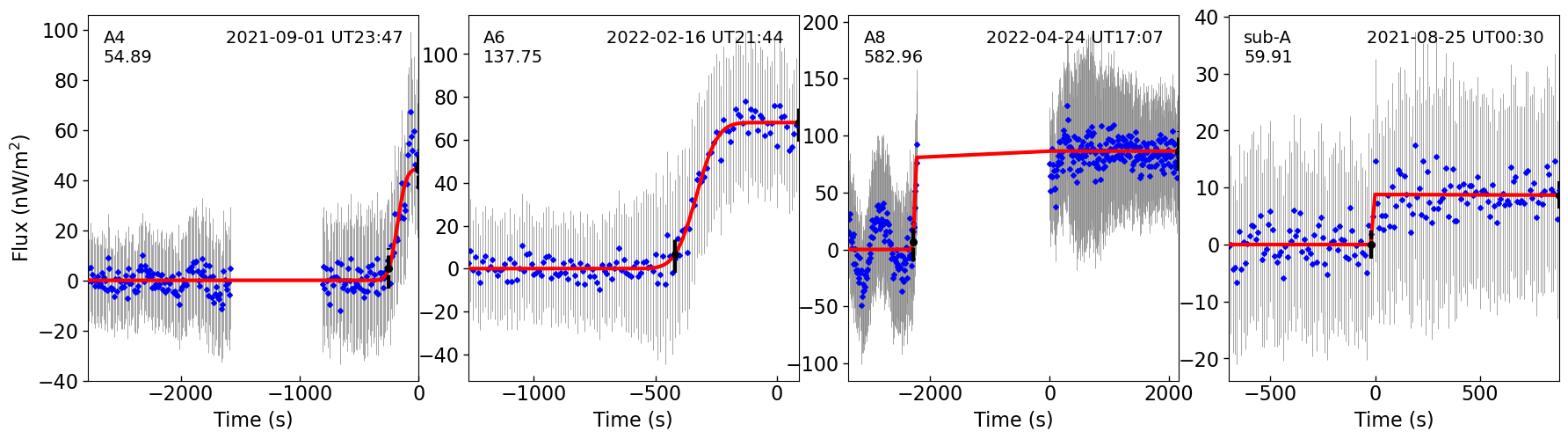}
    \caption{Erroneous flares detected in SD-CH2 due to an apparent rise in background level before the catalog filtering step. {Left:} Non-flare at the end of day {Centre Left:} Residual from curve fit of a larger flare {Centre Right:} Data-taking gap in XSM/CH2. {Right:} Incorrect convergence of one EFP function in a multi-flare group.}
    \label{fig:tauexamples}
\end{figure*}

In Figure \ref{fig:tauexamples}, we see cases when an abrupt change in the background intensity results in a sigmoid-like curve fit. 

\begin{figure}
    \centering
    \includegraphics[width=0.49\linewidth]{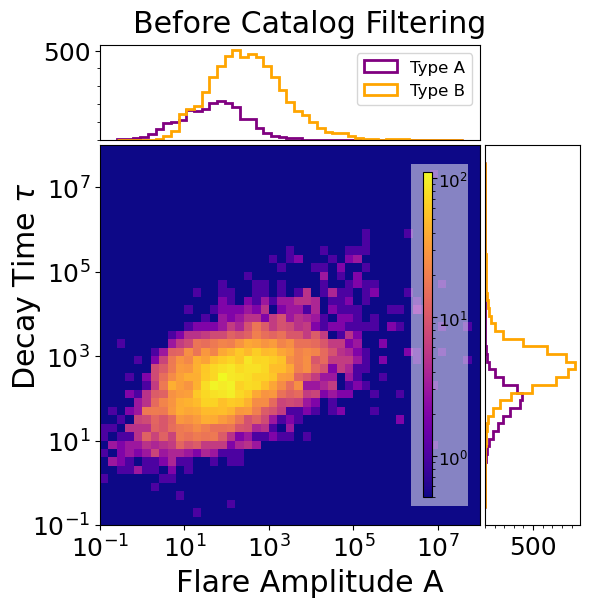}
    \includegraphics[width=0.49\linewidth]{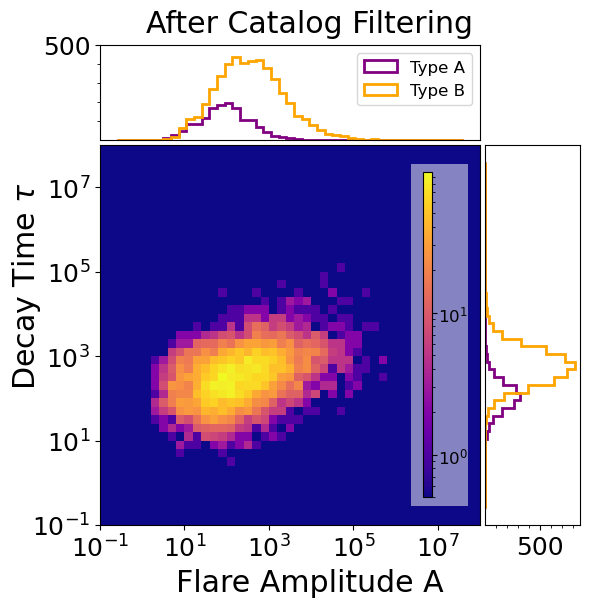}
    \caption{2D histogram of flare amplitude and decay time in log-log scale, of the flares in the flare catalog, before and after the filters are applied.}
    \label{fig:Avstau}
\end{figure}

Given the strong amplitude-decay time dependence of erroneous flares, with most of them have sharp, glitch-like rising phases or drawn out decay phases, a hybrid constraint using the amplitude-decay time 2D histogram can be inferred from the effects of catalog filtering as shown in Figure \ref{fig:Avstau}. We notice that once all the filters, including the reliability filter are placed, the outlier data points are already filtered out. 
In SD-GOES, after applying the flare catalog filters, we are left with 13356 out of the initial 15236 flares, while in SD-CH2, 6266 out of 8203 flares remain.

\section{Monte-Carlo method for power-law curve fitting}\label{sec:montecarlo}

\begin{figure}
    \centering
    \includegraphics[width=0.49\linewidth]{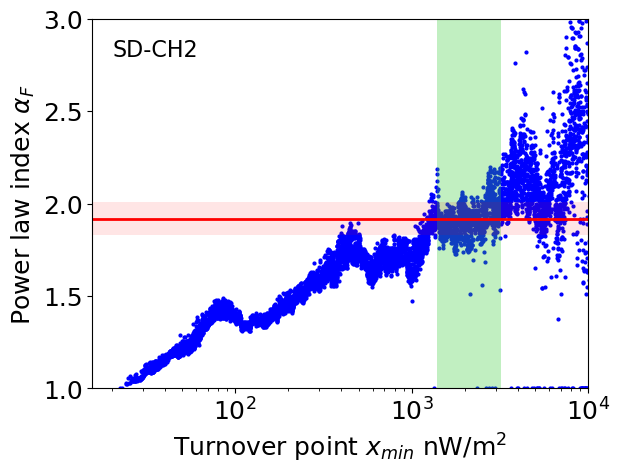}
    \includegraphics[width=0.49\linewidth]{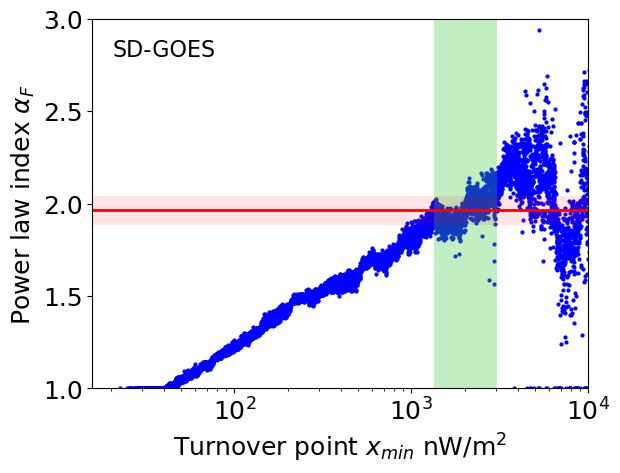}
    \caption{Derived relationship using a Monte-Carlo analysis between power index $\alpha_F$ and turnover point $x_{min}$ for background-subtracted peak flux distributions. The highlighted green region shows the plateau of stability over which the power index is independent of the choice of $x_{min}$. The red line fits this plateau with a constant $\alpha_F$ and the light red highlight is the 1$\sigma$ region.}
    \label{fig:xminvariations}
\end{figure}

{We perform a Monte-Carlo analysis to validate the stability of a power law fit to the peak flux distribution. The distribution is characterised by a single parameter: the number of bins used in the histogram $nbins$. The power law fit is characterised by two parameters: the rightmost bin $x_{high}$, and the leftmost bin $x_{min}$, also referred to as the `turnover point', used in the fit. To infer a valid power index $\alpha_F$ for a power law fit to the histogram distribution, $\alpha_F$ should be stable across a range of values for parameters $nbins$, $x_{min}$ and $x_{high}$.}  

We vary the $nbins$ and the total width of the histogram $(x_{high} - x_{min})$ for various values of the turnover point. $nbins$ is sampled from a uniform distribution [20,50] and $x_{high}$ is sampled from a log-normal distribution with mean $=5$ and standard deviation $=0.5$. The latter distribution is chosen as the number of X class flares $> 10^5$ nW/m$^2$ are low and small number statistics should not drastically affect the fit. We perform a least-squares fit to a power law function $Ax^{-\alpha}$ for each set of parameters over 10,000 samples and plot the value of $\alpha_F$ against $x_{min}$ in Figure \ref{fig:xminvariations}.

The flat region or plateau of stability in Figure \ref{fig:xminvariations} is used to determine the value of the power index. This ranges from $x_{min} \approx 1.4\times 10^3$ to $x_{min} \approx 3 \times 10^3$ nW/m$^2$. We fit an horizontal line to this plateau to obtain $\alpha_F$. This corresponds to $\alpha_F = 1.92\pm 0.09$ and $\alpha_F = 1.96\pm 0.08$ for SD-CH2 and SD-GOES respectively.

%
\begin{acks}\label{sec:acknowledgements}
A.B.V. thanks N.P.S. Mithun for shedding light on the data preprocessing and extraction of level 2 files found on the PRADAN website. We thank Aswin Suresh and Mehul Chandra for help with the background estimation code. We acknowledge the use of data from the Chandrayaan-II, second lunar mission of the Indian Space Research Organisation (ISRO), archived at the Indian Space Science Data Centre (ISSDC). XSM was developed by Physical Research Laboratory (PRL) with support from various ISRO centers. We thank various facilities and the technical teams from all contributing institutes and the Chandrayaan-2 project, mission operations, and ground segment teams for their support.
\end{acks}

%
%
\bibliography{bibliography}
\bibliographystyle{spr-mp-sola}
%
%
%
%

\end{article} 
\end{document}